\documentclass[a4paper,oneside]{revtex4-2}
\usepackage{lipsum}
\usepackage{graphicx}
\usepackage{epsf}
\usepackage{epsfig}
\usepackage{amssymb,amsmath,amsfonts}

\begin{document}
\title{Constraints on a Tidal Charge of the Supermassive Black Hole in M87*  with the EHT Observations
   in April 2017}
\author{Alexander F. Zakharov}
\affiliation{%
$^1$ Bogoliubov Laboratory of Theoretical Physics, Joint Institute for Nuclear Research, 141980 Dubna, Russia; alex.fed.zakharov@gmail.com} 
\affiliation{$^{2}$  MEPhI (Moscow Engineering Physics Institute), National Research Nuclear University, \mbox{Kashirskoe Highway 31}, 115409 Moscow, Russia
} 


\begin{abstract}
More than two years ago the Event Horizon Telescope  collaboration presented the first image reconstruction around the shadow for the supermassive
black hole in M87*. It gives an opportunity to evaluate the shadow size.
Recently, the Event Horizon Telescope collaboration constrained parameters (``charges'') of spherical symmetrical metrics of black holes from an estimated allowed interval for shadow radius from observations of M87* in 2017. Earlier,  analytical expressions for the shadow radius as a function of charge (including a tidal one) in the case of Reissner--Nordstr\"om metric have been obtained.
Some time ago, Bin-Nun proposed to apply a Reissner--Nordstr\"om metric with a tidal charge as an alternative to the Schwarzschild metric in  Sgr~A*.
If we assume that a Reissner--Nordstr\"om black hole with a tidal charge exists in M87*, therefore,
based on results of the shadow size evaluation for M87* done by the Event Horizon Telescope collaboration we constrain a tidal charge. Similarly, we evaluate a tidal charge from shadow size estimates for Sgr~A*.
\end{abstract}


\maketitle

\section{Introduction}

The Very Long Baseline Interferometry (VLBI) technique for observations of compact bright radio sources has been proposed in Union of Soviet Socialist Republics in sixties of the last century and these ideas were realized in the joint experiment United States--Russian experiment proposed by M. Cohen and K. I. Kellermann where 22 m Pushchino and 43~m Green Bank antennas were planned to use, but in the experiment Pushchino antenna was substituted with Simeiz one   \cite{Matveenko_07,Kellermann_92,Lovell_73}.
In eighties Russian astronomers proposed a space--ground interferometer Radioastron which should have an angular resolution at a level of a few microarcseconds at the shortest wavelength 1.3~cm \cite{Kardashev_88,Kardashev_01}.

 Since the best angular resolution was around 8 $\mu$as  for the Radioastron mission\footnote{The space--ground interferometer Radioastron was effectively operating in 2011--2019.} at 1.3~cm or the angular resolution is better than  a visible size of the event horizon for Sgr A*, the Radioastron head  N. S. Kardashev requested to specify general relativistic phenomena, which could be observed, at least in principle, with facilities having such a nice angular resolution and Kardashev and  Langston  expressed the opinion as ``the general relativistic effects due to warped space-time structure near the black hole are directly visible'' \cite{Kardashev_04}.  As a response for the request, in papers \cite{ZNDI05,ZNDI05_2} it was proposed  to treat a dark area in the sky (shadow) as a test for General Relativity (GR) predictions (in spite of the fact, earlier, in many papers and books, see for instance, studies \cite{Bardeen_73,Luminet_79,Luminet_92,Young_76,chandra} where the authors reproduced very similar pictures, however, before our discussions of the issue  people did not claim that the shadow analysis of its size and its shape may be treated as GR test perhaps because  the area of dark region (shadow) is very small). Astronomy is dealing with images, but it is not a common practice to use the image as a test of fundamental physics (e.g., GR) as we proposed in \cite{ZNDI05}.    We  understood that the dark region (shadow) may be reconstructed only with observations of bright structures (mirages)  around the shadows. In 2004--2005 we knew that scattering may spoil a shadow image at 1.3 cm as it was shown in \cite{Falcke_00,Falcke_00_2,Melia_01}\footnote{Based on results of their computer simulations and available estimates of the black hole mass in Sgr A* the authors concluded that the shadow size for the Galactic Center is around 30~$\mu$as \cite{Falcke_00,Falcke_00_2,Melia_01} and it is necessary to use a wavelength around 1~mm to reduce ray scattering and to improve the angular resolution of a
VLBI network. Currently, the predicted shadow size for Sgr A* is slightly more than 50~$\mu$as. An angular resolution of the ground based interferometer Event Horizon Telescope (EHT) is around  25~$\mu$as  \cite{Akiyama_19} and it can not be improved significantly since now the EHT arm lengths are comparable with the Earth diameter.} and   therefore, 1.3 mm (or shorter) is a more suitable wavelength  for a shadow detection as we noted in \cite{ZNDI05}, thus we promoted an opportunity to detect the Sgr A* shadow with
VLBI acting in mm and sub-mm bands, in particular, we discussed an
opportunity to use the ground--space interferometer Millimetron facilities for observations of bright structures around the shadow for its reconstruction in Sgr A* (in these years we did not know about the Event Horizon Telescope (EHT) project).
In 2019 the EHT collaboration presented results of shadow reconstruction for observations of M87* in April 2017 \cite{Akiyama_19}.
The shadow size for M87* is around $42~\mu$as and it  corresponds to black hole mass around $M_{\text{M87*}}=6.5\times 10^9 \,M_\odot$ at distance around $D_{\text{M87*}}=16.8~Mpc$.

{
In last years, theorists proposed a number of alternative theories of gravity including theories with extra dimensions.
In some theories extra dimensions are compactified, in other theories they are  large or even infinite. A review on theories with large and infinite extra dimensions
is presented in \cite{Rubakov_01}.
Around 20 year ago Dadhich et al. showed that if we consider the Randall--Sundrum II braneworld scenario \cite{Randall_1999,Randall_1999b}, the  Reissner--Nordstr\"om metric
may be a black hole solution in the model \cite{Dadhich_00}. The solution is interpreted as a black hole without an electric charge but a tidal charge ($q$)  arising via
gravitational effects from the fifth dimension.
Dadhich et al. \cite{Dadhich_00} proposed to call the corresponding parameter  'tidal charge' since it is arising
from the projection onto the brane of free gravitational field effects in the bulk.
In contrast to the  Reissner--Nordstr\"om metric with an electric charge, where $\mathcal{Q}^2$ is always non-negative,  a tidal charge ($q$)  may be negative.
Later,  Bin-Nun suggested to apply
a Reissner--Nordstr\"om metric with a tidal charge for the black hole at the Galactic Center \cite{Bin_Nun10,Bin_Nun10a,Bin_Nun11}, where a significant
negative tidal charge up to $q=-6.4$  was considered (a definition of $q$ parameter will be given below).
Based on observational constraints of shadow size for Sgr A* given in \cite{Doeleman_08}, we showed  \cite{Zakharov_12}
 that  a significant negative tidal charge should be ruled out with a rather high probability.
In addition, it should be noted that there exist solutions similar to the Reissner--Nordstr\"om metric with a tidal charge in scalar-tensor theories. This approach was proposed by G. Hordeski in \cite{Horndeski_74} and these studies were  forgotten for several years  but a number of applications of Horndeski theories (and their generalizations)
have been considered in last years (see, for instance \cite{Ishak_19}).
In particular, as it was shown  in \cite{Babichev_17} in the framework of the Horndeski approach for scalar-tensor theories a black hole  could have a secondary hair $q$  due to the
non-trivial scalar-tensor mixing and the corresponding solution looks like a Reissner--Nordstr\"om metric with a non-electric charge. Relativistic precession for such  Reissner--Nordstr\"om black holes with a tidal charge are evaluated in \cite{Zakharov_18} (it gives an opportunity to constrain a black hole charge from Very Large Telescope and Keck observations of bright stars moving near the Galactic Center). Such calculations may be useful since the GRAVITY collaboration confirmed
recently a presence of the Schwarzschild precession for the S2 trajectory near the black hole at the Galactic Center \cite{Abuter_20}. Constraints
on parameters of extended theories of gravity have been obtained assuming the congruence with predictions of GR \cite{Borka_21}.
}

{The Equivalence Principle was among the cornerstone assumptions which were used to create GR by A. Einstein. The Weak Equivalence Principle
(WEP) states that two bodies with different compositions and masses fall at the same rate in a gravitational field. Recently, results of data analysis from
the MICROSCOPE space mission were reported and it was shown that there was no violation of WEP at the level around $10^{-15}$ for titanium and platinum pair of materials \cite{Touboul_18,Touboul_19}. Since we used the Reissner--Nordstr\"om metric with a tidal charge and this model looks very similar to the conventional Reissner--Nordstr\"om metric with an electric charge in general relativity we expect that it should be no violation of WEP since from the beginning we suppose that test bodies move along geodesics independently on their masses and chemical compositions assuming that their masses and sizes are small enough. However, only an experiment could indicate that there is a WEP violation or could give additional arguments that there is no WEP violation
in such a problem.
Definitions of Strong  and Weak Equivalence Principles are given in a recent comprehensive review \cite{Tino_20} where also different experimental ways to test the equivalence principle are discussed}.

{In spite of a common opinion that there are supermassive black holes in galactic centers, different alternatives are proposed in last years.
For instance, Ruffini, Arg\"uelles Rueda, ref. \cite{Ruffini_15} proposed to substitute supermassive black holes
with  dense cores from  dark matter. In this scenario dark matter forms  dense cores and  diluted halos in galaxies. In consequent studies, the dark matter distributions were called the RAR-model. Recently, Becerra-Vergara et al. \cite{Becerra_21} declared that in the Galactic Center the RAR model provides a better fit of trajectories of bright stars in comparison with the supermassive black hole model. The properties of bright star trajectories in gravitational field of a dense core in the RAR-model have been considered in \cite{Zakharov_21} and it was concluded that the gravitational field determined in the framework of the RAR model is the harmonic oscillator potential (if the stars move inside a ball with a constant density). In this case trajectories of stars are ellipses with centers at the origin and their periods are the same, so they do not depend on semi-major axes, therefore, it was concluded that these properties are not consistent with existing observational data for trajectories of bright stars. For the Galactic Center
only versions of RAR models are suitable where a dark matter mass around $4.3 \times 10^6~M_\odot$ is inside a ball with a radius less than the smallest pericenter of these stars and total mass of dark matter inside bound trajectories of bright stars (S2, S29, S38 and S55) is less than $3000~M_\odot$ according to \cite{Gravity_21}. In this case the dark matter distributions for these RAR models are also generate the Newtonian potential
as the conventional model of the supermassive black hole. The Newtonian potential of a point like source is a good approximation for the gravitational potential near the Galactic Center  and relativistic corrections and corrections due to a presence of an additional mass in stellar cluster or  in dark matter concentrations may be considered as perturbations of this potential.
If we adopt the RAR model for the galactic centers (for instance, for Sgr A* and M87*), then we cannot expect the formation of shadows in these objects, since photons can freely propagate in galactic cores formed from dark matter and thus, black hole shadows  are not generated. So, since  astronomers observed the shadow for M87* we have to conclude that  RAR model with a dense core instead of the supermassive black hole  is not suitable for M87*. Probably, we have to arrive at the same conclusion in the Sgr A* case}.

{We organized paper in the following way. In Section \ref{sec2} we present motivation of the paper. In Section \ref{sec3} we give analytical expressions for shadow radius as a function of (tidal) charge. Quantitative bounds for M87* and Sgr A* charges are presented in Section \ref{sec4}.  The concluding remarks are given in Section \ref{sec5}.}

\section{Motivation}\label{sec2}
{
Since space plasma is quasi-neutral it is expected that electric charges for astrophysical black holes are negligible (in geometrical units) and
black holes are characterized by only two parameters: its mass and its spin.
The shadow size and shape for Kerr black hole has been considered in \cite{ZNDI05,ZNDI05_2}.  Simplifying results of the paper, we conclude that  a spin slightly changes  shadow sizes while it changes shadow shapes (see, {Figures 2--4} in \cite{ZNDI05}). 
 In~\cite{ZNDI_AA_05} an analytical expression for shadow size as a function of charge is given.
Later, the expression was generalized for a tidal charge case \cite{Zakharov_14}. Below we use the corresponding relation from the paper.
In paper \cite{Kocherlakota_21} the authors quoted \cite{Wald_84} where shadows for Reissner--Nordstr\"{o}m metric were not considered and the authors did not note that there is an analytical expression for the blue curve presented in Figures 1 and {2} given in \cite{Kocherlakota_21}. In the paper we point out that there is an analytical expression for the presented blue curve in Figures {1} and {2} in~\cite{Kocherlakota_21} and we show that our analytical results are consistent with this blue curve. 
An existence of the analytical expression simplifies a comparison of theoretical fits and observational data for Reissner--Nordstr\"{o}m case. We also generalize considerations in
\cite{Kocherlakota_21} for a tidal charge case. Different applications of a tidal charge concept is widely discussed for astrophysical black holes.
A significant negative charge similarly to the values considered in \cite{Bin_Nun10,Bin_Nun10a,Bin_Nun11} increases essentially a shadow size and therefore, if such features would be detected in observations it could be signatures of extra dimension or some other generalizations of GR  similarly to the Horndeski gravity.}

{As it is shown below, the critical impact parameters separating plunge and scatter photon orbits for  the Reissner-Nordström metric may be analytically expressed through a charge parameter. However, as it was noted  significant electric charges are not expected for astrophysical black holes.  The Reissner-Nordström solutions with  tidal charges exist  in the framework of the Randall – Sundrum approach with an extra dimension. At the moment there are no so strict constraints on tidal charges as we have for electric charge constraints. Moreover, a tidal charge could be negative and shadow size would be larger than the shadow size for a Schwarzschild black hole in this case. Therefore, such a shadow excess may be a signature of presence of extra dimension if we adopt the Randall–Sundrum approach}.

\section{An Analytical Expression for Shadow Radius as a Function of Charge}\label{sec3}

In \cite{ZNDI_AA_05} an analytical expression for shadow radius has been obtained as a function of a black hole charge and in the derivation we used
an algebraic condition of vanishing discriminant which was used earlier in \cite{Zakharov_91,Zakharov_94}.
An expression for the Reissner - Nordstr\"{o}m metric \cite{Reissner_16,Nordstrom_18} has the form in natural
units ($G=c=1$) (see, for instance, \cite{MTW} for reference)
\begin {equation}
  ds^{2}=-\left(1-\frac{2M}{r}+\frac{\mathcal{Q}^{2}}{r^{2}}\right)dt^{2}+\left(1-\frac{2M}{r}+\frac{\mathcal{Q}^{2}}{r^{2}}\right)^{-1}dr^{2}+
r^{2}(d{\theta}^{2}+{\sin}^{2}\theta d{\phi}^{2}),
\label{RN_0}
\end {equation}
where $M$ is the mass of a black hole and $\mathcal{Q}$ is its electric charge.
As it is well-known
the motion of a test particle in the $r$-coordinate can be described
by the following equation (see, for example, \cite{MTW})
\begin {eqnarray}
    r^{4}(dr/d\lambda)^{2}=R(r),\label{RN_D 1}
\end {eqnarray}
where $\lambda$ is the affine parameter and
\begin {eqnarray}
&&  R(r)=P^{2}(r)-\Delta({\mu}^{2}r^{2}+L^{2}), \nonumber\\
&&  P(r)=Er^{2}-e\mathcal{Q}r,    \label{RN_D_2}\\
&&  \Delta=r^{2}-2Mr+\mathcal{Q}^{2}. \nonumber
\end {eqnarray}

The constants $ E, \mu, L$ and $e$ are connected with a
particle, i.e., $E$ is energy at infinity, particle,  $\mu$ is its mass, $L$ is
its angular momentum at infinity and $e$ is a particle's charge.
For photon one has $e=0$ and $\mu=0$.
Introducing a new independent variable $\sigma$ as it was done in \cite{Zakharov_94b}  that $d \sigma/ d \lambda = 1/r^2$, one could write equations of motion in a more standard form
\begin {eqnarray}
    (dr/d\sigma)^{2}=R(r),\label{RN_D 1_0}
\end {eqnarray}
where the expression for the polynomial $R(r)$ has the following
form
\begin {eqnarray}
R(r)=E^{2}r^{4}+L^{2}r^{2}+2ML^{2}r-\mathcal{Q}^{2}L^{2}. \label{RN_D_3}
\end {eqnarray}

 Parameters corresponding to  a circular motion in the $r$-coordinate should satisfy to a condition
for the root multiplicity of the polynomial $\hat{R}(\hat{r})$ (in this case at this multiple root one has $dr/ d \sigma = d^2 r/ d \sigma^2=0$)
and introducing dimensionless variables one obtains
\begin {eqnarray}
\hat{R}(\hat{r})={R(r)}/({M^{4}E^{2}})={\hat{r}}^{4}-\xi^{2}{\hat{r}}^{2}+2{\xi}^{2}\hat{r}
 -{\hat{\mathcal{Q}}}^{2}{\xi}^{2}.\label{RN_D_4}
\end {eqnarray}
where $\hat {r}=r/M, \xi=L/(ME)$ and $\hat
{\mathcal{Q}}=\mathcal{Q}/M.$ Below we omit the hat symbol
for these quantities.
Introducing the notations $l=\xi^{2}, q=\mathcal{Q}^{2}$, we obtain
\begin {eqnarray}
    R(r)=r^{4}-lr^{2}+2lr-ql.\label{RN_D_5}
\end {eqnarray}

The polynomial $R(r)$ thus has a multiple root  if and only if the polynomial discriminant is vanishing \cite{Kostrikin_82}
and for the polynomial in Equation (\ref{RN_D_5}) as it was shown in \cite{Zakharov_14} one has
\begin {eqnarray}
  l^{3}[l^{2}(1-q)+l(-8q^{2}+36q-27)-16q^{3}]=0. \label{RN_D_7}
\end {eqnarray}

For the case $l=0$ one has a multiple root at
$r=0$, we find
 that the polynomial $R(r)$ has a multiple root for $ r\geq r_{+}$ if and only if
\begin {eqnarray}
 l^{2}(1-q)+l(-8q^{2}+36q-27)-16q^{3}=0. \label{RN_D_8}
\end {eqnarray}

If $q=0$, we obtain the well-known result for a Schwarzschild black
hole \citep{MTW}, $l_{\rm cr}=27$, or
$\xi_{cr}=3\sqrt{3}$ (where $l_{\rm cr}$ is the positive root of Equation
(\ref{RN_D_8})). If $q=1$, then $l = 16$, or \mbox{$\xi_{cr}=4$ \cite{ZNDI_AA_05,Zakharov_14}}, which
also corresponds to results presented in Figure 2 in
paper \cite{Kocherlakota_21}. Solving Equation (\ref{RN_D_8}), one has
   \begin {eqnarray}
l_{\rm cr}=\frac{(8q^{2}-36q+27)+\sqrt{D_1}}{2(1-q)}, \label{RN_D_9}
\end {eqnarray}
where
$D_1=(8q^{2}-36q+27)^{2}+64q^{3}(1-q)=-512\left(q-\dfrac{9}{8}\right)^3.$
It is clear from the last relation that there are circular unstable
photon orbits only for $q \le \dfrac{9}{8}$.
For $1< q \le \dfrac{9}{8}$ one has unstable photon orbits but there are no shadows (the naked singularities forming shadows were considered in \cite{Shaikh_18}).
 The photon capture cross section for a charged
    black hole  turns out to be considerably smaller than the capture cross section of a
 Schwarzschild black hole as one can see in corresponding figures presented in \cite{ZNDI_AA_05,Zakharov_14,Kocherlakota_21}. The critical value of the impact parameter,
 characterizing the capture cross section for a Reissner--Nordsr\"om black hole, is determined by the Equation (\ref{RN_D_9}), since $\xi=\sqrt{l}$.
Substituting
Equation (\ref{RN_D_9}) into the expression for the coefficients of the
polynomial $R(r)$ it is easy to calculate the radius of the unstable
circular photon orbit (which is the same as
 the minimum periastron distance). The orbit of a
 photon moving from infinity with the critical impact parameter, determined
in accordance with Equation (\ref{RN_D_9}) spirals into  circular orbit.
To find a radius of photon unstable orbit we will solve equation substituting $l_{\rm cr}$ in the relation
\begin {eqnarray}
 \dfrac{\partial{R}}{\partial{r}} =
  2\left( 2r^{3}-l_{\rm cr}r+l_{\rm cr}\right)=0.\label{RN_D_5_2}
\end {eqnarray}

From
trigonometric formula for roots of cubic equation we have
\begin {eqnarray}
r_{\rm crit}=2\sqrt{\frac{l_{\rm cr}}{6}} \cos{\frac{\alpha}{3}},
\label{RN_D_10}
\end {eqnarray}
where
\begin {eqnarray}
\cos \alpha={-\sqrt{\frac{27}{2 l_{\rm cr}}}}, \label{RN_D_11}
\end {eqnarray}

\vspace{-9pt}
\begin{figure}[ht!]

\includegraphics[width=0.75\textwidth]{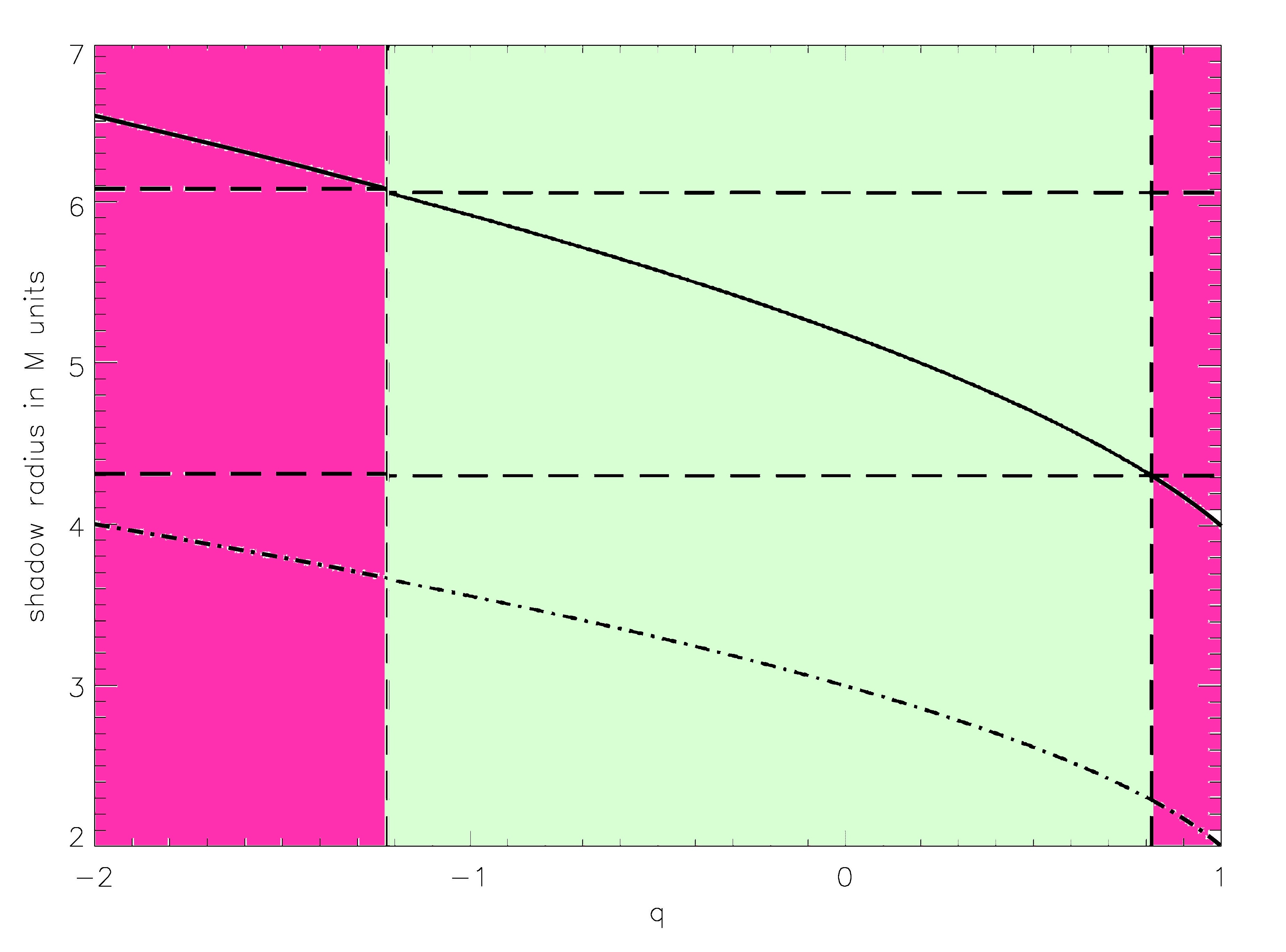}

\caption{Shadow (mirage) radius (solid line) and radius of the last
circular unstable photon orbit (dot-dashed line) in $M$ units as a
function of
 $q$. Similarly {to} \cite{Kocherlakota_21}
 we adopt $\theta_{\text{sh~M87*}} \approx 3\sqrt{3}(1
\pm 0.17)\,\theta_{\text{g~M87*}}$, at 68\% confidence levels as it was given {in}
\cite{Psaltis_20}. Horizontal dashed lines correspond to constraints on shadow radius in $M$ units, namely, $r_{low}=4.31$ and $r_{upper}=6.08$.
{Light green vertical strip corresponds to $q$ parameters which are currently consistent with the shadow size estimate done by the EHT {collaboration} \cite{Kocherlakota_21} while rose strips correspond to $q$ parameters which are not consistent with the shadow size estimate}.}
 \label{Fig1}
\end{figure}

\section{Constraints on a Tidal Charge}\label{sec4}

\subsection{Constraints from Observations of M87*}
Similarly to  \cite{Kocherlakota_21} we adopt $\theta_{\text{sh~M87*}} \approx 3\sqrt{3}(1
\pm 0.17)\,\theta_{\text{g~M87*}}$, at 68\% confidence levels, or $\theta_{\text{sh~M87*}} \in [4.31, 6.08] \theta_{\text{g~M87*}}$, where $\theta_{\text{g}~M87*} \approx 8.1~\mu$as  since $\theta_{\text{g~M87*}}=2M_{M87*}/D_{M87*}$, one obtained
$q \in [-1.22, 0.814]$ from Equation (\ref{RN_D_9}), see also Figure \ref{Fig1} and the upper bound ($q_{upp}=0.814$) of the interval corresponds   to the upper limit $\mathcal{Q}_{upp}=\sqrt{q_{upp}} \approx  0.902$ which is approximately equal to the value shown in Figure 2 in \cite{Kocherlakota_21}.

\subsection{Constraints from Observations of Sgr A*}

 Doeleman et al. \cite{Doeleman_08} gave preliminary estimates of ring sizes, namely, an inner ring was estimated around $35~\mu$as and an outer ring was around $80~\mu$as. Therefore,
a shadow size should be in the interval  $\theta_{\text{sh~Sgr~A*}} \in [3.5, 8.0] \theta_{\text{g~Sgr~A*}}$ ($\theta_{\text{g~Sgr~A*}} \approx 10~\mu$as for Sgr A*). From these estimates one could conclude that $q > -5.26$ from Equation (\ref{RN_D_9}) taking into account that $\theta_{\text{sh~Sgr~A*}} < 80~\mu$as.

Later, the EHT collaboration presented different estimates of using different models~\cite{Lu_18}, for instance, for Model B (a ring model) it was given for S1 size around $(52\pm 2) \mu$as and for S2 size around $(25\pm 2) \mu$as, therefore, similarly to the M87* case one has $\theta_{\text{sh~Sgr~A*}} \in [2.80, 5.4] \theta_{\text{g~Sgr~A*}}$, therefore, we obtain that $q> -0.25$ taking into account the upper limit of the interval $\theta_{\text{sh~Sgr~A*}}$, while the lower limit does not constrain $q$ parameter since
 the minimal shadow size for $q=1$ is  $\theta_{\text{sh~Sgr~A*}} = 4~\theta_{\text{g~Sgr~A*}}$ for a Reissner--Nordstr\"om metric.
 Astronomical community is waiting for further clarifications of a shadow ring size for \mbox{Sgr A*} with new observational data.

\section{Discussion}\label{sec5}

{In order to test as an opportunity of the Reissner-Nordström metric with a tidal (or electric) charge for the M87* or Sgr A*
it is necessary to decrease an allowed interval for shadow radii or in other words to improve an accuracy of shadow size estimates.  Currently, observational data are consistent with negative, vanishing and positive $q $ parameter. If in future a shadow size estimate will be significantly larger than it is expected in the Schwarzschild black hole case (or, in other words, the Schwarzschild photon ring radius is less
than the left boundary of the confidence interval for allowed shadow radii evaluated from observations)
  then we will  conclude that  the Reissner-Nordström metric with a tidal charge might be considered as a more suitable model instead of the Schwarzschild black hole. In the opposite case, if  a shadow size estimate will be significantly smaller than it is expected in the Schwarzschild black hole case  then the Reissner-Nordström metric might be considered as a suitable model. If in future an allowed internal for shadow radii will be significantly decreased and a shadow size will be still consistent with the Schwarzschild black hole model we will  conclude that probably a consideration of black hole charge is not needed to fit observational data in this case.}

Many years ago J. A. Wheeler claimed that black holes completely determined by only three parameters: mass, electric charge and angular momentum (the statement is also called ‘no hair theorem’). In this case it was assumed that the metric must be asymptotically flat and, therefore, the Kerr--Newman solution is the general case of black hole metric with an electric charge while the Reissner--Nordstr\"om metric is the general case of spherically symmetric metric. Really,  the Reissner--Nordstr\"om metric is the general spherically symmetric solution of Einstein--Maxwell equations. The Birkhoff theorem states that in GR any spherically symmetric solution of the vacuum field equations must be given by the Schwarzschild metric and for spherically symmetric matter distribution an exterior metric is also  Schwarzschild one  \cite{MTW,DInverno_98,Deser_05}. There are no contradictions with the Birkhoff theorem in considerations of Reissner--Nordstr\"om metric with a electric charge, a tidal charge or ``charge'' due to interactions with  scalar fields as it was done in Horndeski approach \cite{Babichev_17} because different problems are considered in these cases. Assuming spherical symmetry for all cases in the case of the Birkhoff theorem we search for the general vacuum solution of Einstein equations, in a derivation of the Reissner--Nordstr\"om metric with an electric charge we search for the general vacuum solution of Einstein--Maxwell equations, in a derivation of the Reissner--Nordstr\"om metric with a tidal
charge Dadhich et  al. \cite{Dadhich_00} considered free gravitational field effects in the bulk and their projection onto the brane, in the case
of  Horndeski approach a simple solution was found which looked similar to Reissner--Nordstr\"om one but its parameter is not an electric charge but it characterises
a scalar-tensor interaction.
 In the considerations of different problems different assumptions have been chosen and as a result different solutions for listed problems have been found.

In spite of the fact that the Reissner--Nordstr\"om metric is a rather general solution, a range of its applications in astronomy
looks very limited since in macroscopical volumes plasma is usually quasi-neutral and it is natural for astrophysical black holes to expect  very small electric charges in comparison with their masses. For instance, in \cite{Zajacek_19} the authors found that the maximal charge of the supermassive black hole at the Galactic Center is still 12 orders of magnitude smaller than the extremal charge of Sgr A*. Since deviations of the Reissner--Nordstr\"om metric from the Schwarzschild metric are rather small for such small electric charges, impact of electric charges on trajectories of particles and photons are practically
negligible. In the case of a tidal charge and an appearance of similar parameter in the scalar-tensor Horndeski approach  (as it was shown in \cite{Babichev_17}) we have no so strict constraints on $q$ parameter and initially we could assume a significant negative $q$ parameter, after that we can test and evaluate a tidal charge in different astronomical systems.

Solar system constraints on a tidal charge were considered in \cite{Bohmer_2008}, where the authors analysed the deflection of light, radar echo delay and
and the perihelion precession of Mercury orbit.
However, the observational data for the perihelion precession and deflection of light are affected by poorly known higher order harmonics in gravitational potential of Sun and it is not easy to catch small deviations from Schwarzschild field due to a possible presence of a tidal charge. Soon after M87* shadow reconstruction by the EHT collaboration~\cite{Akiyama_19}, estimates of spin and tidal charge were done in \cite{Banerjee_20,Neves_20}. In addition, tidal charge estimates for different astronomical objects are given in \cite{Neves_20}.

\section{Conclusions}

Since an angular resolution 8~$\mu$as of the Radioastron interferometer at the shortest wavelength 1.3~cm is suitable for shadow reconstruction,
in papers \cite{ZNDI05,ZNDI_AA_05} we suggested to use these facilities for shadow observations in Sgr A*, but we also noted that VLBI observations in mm band are much more suitable since as it was in \cite{Falcke_00} scattering at cm wavelengths did not give an opportunity to reconstruct shadow shapes (see, also results of observations in \cite{Johnson_18,Issaoun_19,Issaoun_21}).
In papers \cite{ZNDI05,ZNDI_AA_05} we noted that if shadows could be reconstructed, black hole parameters, in particular its charge and spin could be evaluated. In \cite{Akiyama_19} the first shadow reconstruction for M87* was given. Using constraints on shadow radius similarly to \cite{Kocherlakota_21}, we conclude that for a tidal charge in M87* one has $q >  -1.22$.  In \cite{Bin_Nun10,Bin_Nun10a,Bin_Nun11}  Bin-Nun suggested to use
the Reissner--Nordstr\"om metric with a tidal charge as a black hole model for Sgr A*. As we showed based on constraints on shadow size with EHT observations, a significant negative tidal charge is not probable for Sgr A* since for this black hole we obtained $q>-0.25$. Therefore,  we constrained  tidal charges for M87* and Sqr A*
   using an analytical expression of a shadow size  as a function of a tidal charge found in \cite{Zakharov_14}, see  Equation (\ref{RN_D_9}).

        Summarizing, we could claim that currently we can not exclude negative $q$ parameter for both Sgr A* and M87*, therefore, based on EHT collaboration observations we can not reject  Reissner--Nordstr\"om metrics with  tidal charges for these objects and we need further observations to confirm or disprove a presence of extra dimensions considered in the framework of the Randall--Sundrum approach.

 { As it was noted earlier,  it is necessary to improve shadow size estimates in  future since currently a shadow size is comparable with the accuracy of the EHT facilities. Therefore, an incorporation of antennas operating at 0.6~mm wavelength as it is planned in the new generation of the EHT project (ngEHT) is very useful. Another  way to reduce significantly a range for allowed shadow radii is a creation of ground – space interferometer like Millimetron or similar facilities acting in mm and sub-mm bands.
    }

\acknowledgments{It is a pleasure to thank
  Michael Johnson, Luciano Rezzolla, Rajibul Shaikh and Maciek Wielgus for useful
discussions on shadows reconstructed from M87* observations with EHT team and constraints on charges for Reissner--Nordstr\"om  black holes.
The author appreciates also  referees for constructive critical remarks.}





%





\end{document}